# Organic nanodiamonds

Todd Zapata,[1] Neil Bennett,[2] Viktor Struzhkin,[2] Yingwei Fei,[2] Fedor Jelezko,[3] Johannes Biskupek,[4] Ute Kaiser,[4] Rolf Reuter,[5] Joerg Wrachtrup,[5] Fahad Al Ghannam,[6,7,8] and Philip Hemmer[1,6]

[1] Electrical & Computer Engineering Department (ECEN), Texas A&M University, College Station, TX 77843, USA
[2] Geophysical Laboratory, Carnegie Institution of Washington, 5251 Broad Branch Road, Washington DC 20015-1305, USA
[3] Institut für Quantenoptik, Universität Ulm, Albert-Einstein-Allee 11, D-89081 Ulm, Germany
[4] Zentrale Einrichtung Elektronenmikroskopie, Arbeitsgruppe Materialwissenschaftliche Elektronenmikroskopie, Universität Ulm, Albert-Einstein-Allee 11, D-89081 Ulm, Germany
[5] Univ of Stuttgart, 3. Physikalisches Institut, Universität Stuttgart, Pfaffenwaldring 57, 70569 Stuttgart, Deutschland
[6] The Institute for Quantum Science and Engineering (IQSE), Texas A&M University, College Station, TX 77843, USA
[7] Physics Department, Texas A&M University, College Station, TX 77843, USA
[8] The National Center for Applied Physics, KACST, P.O.Box 6086, Riyadh 11442, Saudi Arabia

Abstract: Nano-crystalline diamond is a new carbon phase with numerous intriguing physical and chemical properties and applications. Small doped nanodiamonds for example do find increased use as novel quantum markers in biomedical applications. However, growing doped nanodiamonds below sizes of 5 nm with controlled composition has been elusive so far. Here we grow nanodiamonds under conditions where diamond-like organic seed molecules do not decompose. This is a key first step toward engineered growth of fluorescent nanodiamonds wherein a custom designed seed molecule can be incorporated at the center of a nanodiamond. By substituting atoms at particular locations in the seed molecule it will be possible to achieve complex multi-atom diamond color centers or even to engineer complete nitrogen-vacancy (NV) quantum registers. Other benefits include the potential to grow ultrasmall nanodiamonds, wherein each diamond no matter how small can have at least one bright and photostable fluorescent emitter.

From the beginning of artificial diamond growth, nanodiamonds were grown from organic compounds[1]. However due to the emphasis on growing large diamonds, this early work was largely forgotten. Recently, the emergence of certain fluorescent color centers in diamond has rekindled interest [2] because of the need for better control over nanodiamond properties. For example, ultra-small high-quality nanodiamonds with nitrogen-vacancy (NV) centers are needed for future nanoscale magnetic sensing applications [3]. In addition, more deterministic growth techniques are needed to fabricate large quantities of ultrasmall nanodiamonds with near-unity yield of stable fluorescent emitters like the silicon-vacancy (SiV) [4] or nickel-nitrogen centers (ex: NE8) [5]. Unfortunately the high pressure high temperature (HPHT) growth of nanodiamonds starting from organic material has always required temperatures that were so high that the initial molecules were completely decomposed before the onset of diamond growth[6]. This makes a bottom-up engineering approach to fluorescent nanodiamond fabrication impractical. Here we overcome this key problem, dramatically lowering the diamond-growth temperature to well below the decomposition temperature of proposed organic seed molecules. Under these low-temperature conditions we succeeded in growing nanodiamonds in the range of 2-100 nm.

The basic idea of seeded nanodiamond growth is illustrated in Figure 1a. An organic seed molecule containing one or more diamond lattice units is synthesized chemically. Dopant atoms are incorporated into this seed molecule in a geometry that will enable conversion into a specific color center of interest either during or after the nanodiamond growth. A source of reactive carbon (ex: methyl and ethyl radicals) is then added to the growth mix. This is supplied by a hydrocarbon that "cracks" at a lower temperature than the diamond-like seed molecule. At low concentrations (below the self-nucleation threshold) the carbon radicals should add one at a time to the seed molecule to slowly grow high-quality diamond around the seed. Methyl and ethyl carbon radicals are shown in Figure 1 because these are the dominant radicals present during plasma-based chemical vapor deposition (CVD) growth, but at high pressure other reactive carbon species, like halogenated carbon compounds may also be used [7]. Here we note that a similar diamondoid seeding approach has successfully been applied to enhance nucleation of nanodiamonds for the growth of superior nanocrystal CVD diamond films [8] [9]. Other prior work has shown that isobutene molecules can also serve as a source of reactive carbon in the synthesis of larger diamondoids from smaller ones [9]. Larger diamondoids have also been grown from smaller ones in xenon plasma.[10]

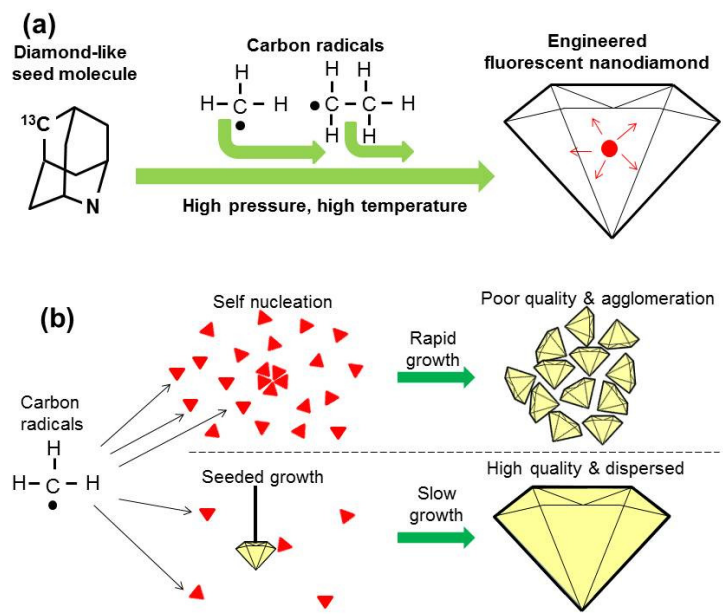

Figure 1. The basic concept of engineered fluorescent nanodiamonds via molecule seeded growth. (a) A diamond-like seed molecule is chosen that has specific atoms arranged in the approximate locations needed to form some color center of interest (the example shown is aza-adamantane with a 13C carbon that could be a precursor for a nitrogen-vacancy quantum register,[11] which requires a well-defined spacing between a NV center and a single 13C in an otherwise pure 12C diamond). Reactive carbon (like the methyl and ethyl radicals shown) is then created by cracking a hydrocarbon that decomposes at a much lower temperature than the diamondoid seed molecule. The subsequent growth of a diamond around the seed molecule could give near-deterministic placement of the desired color center, and assure at least one fluorescent emitter per nanodiamond no matter how small. (b) Conventional nanodiamond growth from organic precursors relies on decomposition followed by self-nucleation. For this to occur, the concentration of reactive carbon (radicals) must be high enough that

**multiple radicals spontaneously nucleate with a high probability. However once nucleation has occurred, the high radical concentration leads to subsequent rapid growth which tends to produce lower quality diamond. In contrast, for seeded growth the reactive carbon concentration can be kept much lower so that diamond growth can be more controlled and therefore higher crystal quality and chemical purity might be possible.**

Growing nanodiamonds slowly, around a diamond-like seed molecule, could potentially produce higher quality diamond. For example, slow growth has been used for the deterministic orientation of NV centers, [12] [13] [14] and the successful n-type doping of diamond with phosphorous. [12] [15] In contrast, self-nucleation growth of nanodiamond requires a relatively high reactive carbon concentration which in turn leads to rapid subsequent growth that can produce lower quality diamond with more defects, as illustrated in Figure 1(b)[16].

Ideally the seeded diamond growth process would encapsulate the seed molecule so that it ends up near the center of the nanodiamond allowing near-deterministic creation and placement of the desired color center. Even a complex multi-atom color center might be produced by this technique with a much higher yield than relying on the probabilistic co-location of the necessary atoms. In addition the diamond grown around the seed can have a very pure chemical and/or isotopic composition, since doping of the reactive carbon source is not required. This is important for color centers like the NV whose magnetic sensitivity is quickly degraded by the presence of other nitrogen atoms or other spin impurities in the diamond lattice [17]. Similarly a quantum register consisting of a NV center optical-spin interface and a 13C storage nucleus placed at an optimum distance could be grown into an isotopically pure 12C diamond to give maximal storage times [18].

The key first step on the path toward seeded nanodiamond growth is to demonstrate that diamond can be grown under conditions where the organic seed molecules do not appreciably decompose. This is most easily done by growing diamond directly from a diamondoid molecule which is a single lattice unit of diamond; namely adamantane (Sigma-Aldrich 98%), as shown in Figure 2. Clearly since adamantane is the only molecule present, some decomposition is needed to produce enough reactive carbon for diamonds to grow, but this should not consume a significant amount of the starting material. Experimentally, the stability of the starting material is verified by monitoring the Raman spectra of the C-C and C-H stretch regions of adamantane during a nanodiamond growth run in a diamond anvil cell (DAC), as shown in Figure 2(a). More details of the benefits and tradeoffs of a DAC are discussed in Figure 4. Although there are changes in the Raman spectra during growth, the final post-growth Raman spectra look the same as those seen at the much lower temperature of 317 C where the pressures are nearly the same. Thus, no significant chemical decomposition occurred. Note that these spectra do not rule out some conversion to higher diamondoids as the C-C and C-H bands are broad at high pressure. For comparison, Figure 2(c) shows clear evidence of adamantane decomposition at higher growth temperatures that can be accessed by laser heating. Here the C-H stretch, which is a sensitive measure of decomposition [19] for adamantane, significantly distorts during this growth run. In fact, this distortion is accompanied by visible white flashes.

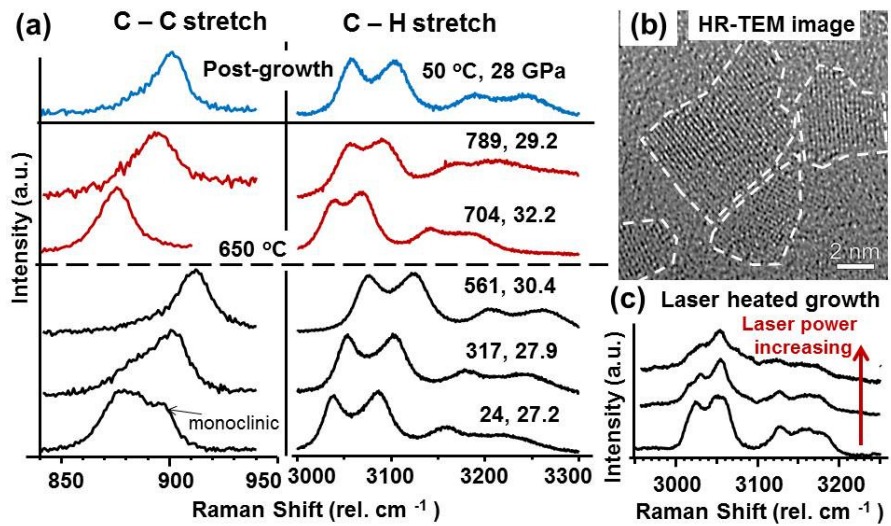

Figure 2. Proof of the chemical stability of an organic seed molecule (adamantane) under conditions where diamond can grow. (a) Adamantane Raman spectra during diamond growth. Temperature and pressure are as indicated. As the temperature begins to increase both the C – C and C – H stretch frequencies gradually increase by an amount that is consistent with the increasing pressure. The C – C band also shows the gradual transition from tetragonal to monoclinic phase which occurs in this pressure range [20]. However at 650 C (horizontal dashed line) there is a sudden red-shift of both spectra along with an increase in pressure, possibly signaling a phase change. Further increases in temperature coincide with a drop in pressure suggesting that some fraction of the material is reacting or decomposing. At the same time the Raman lines broaden but continue to shift toward larger wavenumbers even though the pressure is now dropping. However once the diamond growth run is completed and the temperature is allowed to return to ambient (post-growth), both bands return to peak positions and linewidths that are consistent with the Raman spectra of acquired at much lower temperature (~317 C trace) whose pressure matches the post-growth pressure. (b) HR-TEM image of nanodiamonds produced in the adamantane-only growth run. The dominant products were nanodiamonds in the range of 2-10 nm and carbon onions in the range of 10-50 nm. (c) For comparison, a much higher-temperature growth run which employed laser heating shows significant distortion of the Raman C – H band. Such spectral distortions are common when decomposition and/or reactions are taking place [19]. In addition, white flashes were seen at the highest growth temperatures which is further evidence of decomposition[21]. Unfortunately the actual temperature could not be measured by our thermal imaging apparatus because it was below the minimum temperature of 1200 C.

Proof that nanodiamonds can be grown under conditions where the adamantane does not appreciably decompose is given in Figure 2(b). Here a high resolution transmission electron microscope (HR-TEM) image shows nanodiamonds isolated from the reaction products. The diamond size range is 2-10 nm. It should be noted that smaller diamonds might also be present but cannot be clearly imaged through the background of the thin amorphous carbon film (ca. 10 nm thick) that acts as sample support. In agreement with prior nanodiamond growth at much higher temperatures, other forms of nanocarbon were also found, in this case carbon onions larger than 10 nm [6].

Having established that nanodiamonds can grow under conditions where diamondoid seed molecules remain stable, the next step is to lower the growth temperature even further by growing from a mixture of diamond-like seeds plus an easily cracked source of reactive carbon. This is demonstrated in Figure 3(a) using paraffin (Sigma-Aldrich) as a source of reactive carbon, and adamantane as seed molecules. Here the HR-TEM image in Figure 3(a) shows a representative 8 nm diamond grown at an even lower temperature than in pure adamantane (Figure 2). Diamonds ranging in size from a few nanometers up to 10's of nanometers were observed in this growth run. Paraffin was chosen as the reactive carbon source because it can be loaded in the DAC as a solid at room temperature, and cracks at a relatively low temperature.[22] During growth, only the paraffin C-H stretch could be seen in the Raman spectra (not shown) and it eventually disappeared as the growth temperatures increased indicating that material was being consumed. At the highest growth temperature (635 C) a G-peak indicative of graphite was seen, and when the product is removed from the DAC a D-peak was also visible, which is associated with either amorphous carbon or small nanodiamonds.[23] In addition to paraffin, diamond was grown from other organic molecules. In particular, Figure 3(b) shows a large round diamond that was grown from heptamethylnonane seeded with 1-adamantylamine. Here the growth temperature was around 550 C after an initial spike to 600 C for 90 seconds.

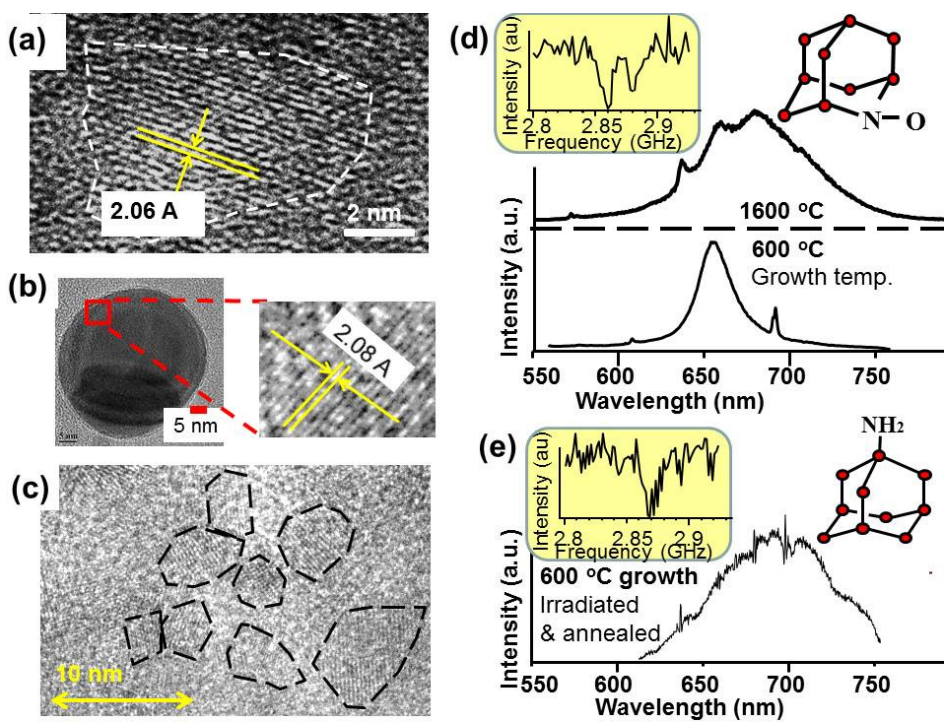

**Figure 3.** Experimental demonstration of diamond growth seeded with organic molecules. (a) HR-TEM image of nanodiamonds found in the product of adamantane seeded growth from paraffin. The magnified nanodiamond has a lattice plane spacing of about 2.06 Å which is close to the 111 lattice plane spacing in bulk diamond. Nanocrystals of graphite were also found (not shown) but significantly no carbon onions. (b) HR-TEM image of a nanodiamond produced by 1-adamantylamine seeded heptamethylnonane growth. Again the lattice plane spacing is close to the 111 spacing in bulk diamond. (c) TEM image of nanodiamonds grown using a multi-anvil cell with paraffin wax as the reactive carbon source and 1-adamantylamine as the seed. The growth run was longer than all others (~ 24 hrs). The image has nanodiamonds with sizes ranging from 2 to 10 nm but diamonds as large as 100 nm were seen in other locations. The electron diffraction pattern (not shown) includes cubic diamond 111 lattice reflections. (d) Post-growth spectra showing that different color centers are produced for different growth temperatures as indicated. The top trace corresponding to a 1600 C growth temperature shows the characteristic nitrogen-vacancy (NV) fluorescence after annealing the product in air at 700 C. Inset shows the corresponding optically detected magnetic resonance (ODMR) spectrum. The growth mix consisted of adamantane and 2-Azaadamantane *N*-Oxyl (AZADO). The bottom trace corresponds to a 600 C growth temperature for AZADO seeded paraffin and shows an unknown color center that appears after a 10 minute anneal in air at 530 C. Sharp peaks in the spectra at 694 nm are ruby. (e) It was possible to observe nitrogen-vacancy color centers in product extracted from a low temperature growth run, but only after electron irradiation followed by and vacuum annealing. Here the starting material was paraffin seeded with 1-adamantylamine and growth lasted for 90 min at 550 C after an initial 90 sec at 600 C. Initially, no color centers were found in the product of this growth run. However after irradiation, strong fluorescence appears whose optical spectrum matches that of NV in nanodiamonds (the spikes are spectrometer noise). Inset shows the corresponding ODMR.

The effect of growth temperature on the fluorescent color centers produced during diamond growth is presented in Figure 3(d). As seen, growth temperature (which controls growth rate) is found to strongly influence which diamond color centers are produced even for the same dopant molecules in the growth mix. In the top trace of Figure 3(d) a high temperature growth with a nitrogen (and oxygen) containing seed molecule, 2-Azaadamantane *N*-Oxyl (AZADO, Sigma-Aldrich), produced numerous nitrogen-vacancy (NV) centers after baking the product in air at 700 C. Notably, no irradiation was required to produce these NV centers implying that defects (vacancies) were incorporated into the diamond crystal during the rapid growth. However at lower growth temperature no NV is produced. Instead an unknown color center is observed (after 10 minutes baking in air at 530 C), as shown in the bottom trace of Figure 3(d). The fact that no NVs formed during low temperature growth, even though nitrogen was present in the seed molecule, means that defects like vacancies were excluded from the lattice. This may be interpreted as indirect evidence that slower growth produces higher quality diamond.

To verify that nitrogen is in fact incorporated into the nanodiamonds during low temperature growth, post-irradiation was performed to produce NV centers, as illustrated in Figure 3(e). For this data the diamonds were grown from paraffin seeded with 1-adamantylamine (Sigma Aldrich). The growth was done at an even lower temperature (550 C) but for a longer time (90 min) than in the AZADO seeded runs. Again nanodiamonds of various sizes were found in the HR-TEM data, but initially no NV spectra. Subsequent electron irradiation of the nanodiamonds followed by vacuum annealing at 700 C for 2 hours did produce NV. The irradiation was performed in a JEOL 2010 TEM at an energy of 200 kV. The resulting emission spectrum in Figure 3(e) is consistent with NV emission in nanocrystals. [24] We also observed optically detected magnetic resonance (ODMR) in these nano-diamonds (inset of Figure 3(e)) which is a signature of the NV center.

So far we have not able to prove that the seed growth illustrated in Figure 1(a) is occurring. However, a control experiment was performed using paraffin without diamondoid seed molecules. In this case no solid product remained after baking in air above 530 C for 10 minutes, which was the temperature needed to activate nanodiamond fluorescence in the case of nitrogen-doped diamondoid seeds.

To test the feasibility of "scaling up" the process, nanodiamonds were grown inside a multi-anvil press for 24 hours at 8 GPa and 560 C, Figure 3(c) Details about the multi-anvil press experimental technique can be found in Supplementary Information and [25]. The reactive carbon source was paraffin wax and the seed was 1-adamantylamine. Nanodiamonds with a size range of 5 to 100 nm were seen and the electron diffraction pattern (not shown) included the characteristic diamond (111) reflection. Nitrogen-vacancies were also observed after irradiation (not shown).

In conclusion we demonstrated that diamond can be grown under conditions were organic seed molecules are stable. This a key requirement for future growth of seeded nanodiamonds. While we could not confirm seeded growth in our experiments, no diamonds were found in control experiments without seed molecules present. If successful, the molecule-seeded approach to nanodiamond growth promises to give an unprecedented level of control over nanodiamond properties and color centers therein. This has numerous potential applications ranging from ultra-small non-bleaching fluorescent biological probes (down to sizes smaller than dye molecules), to nanoscale NV-based sensors of magnetic and electric fields [26] and temperature [27], to precisely engineered quantum registers with deterministic placement of nuclear storage qubits and NV optical interface qubits.

Future work mainly centers on demonstrating and optimizing the seeding process. One way to establish seeded growth is to grow nanodiamonds with a deterministic number of fluorescent centers like NV, or SiV. Another way is to use a different isotopic composition for seed and growth medium, for example a 13C seed doped with nitrogen and natural abundance paraffin. In this case the NV color centers produced would have a distinct 13C signature if the seeding were successful. Optimization of the seeding process primarily consists of enhancing the fraction of diamondoid seeds that produce nanodiamonds. One approach is to identify catalysts, for example halogenated carbon was shown to catalyze diamond growth in prior work.[7] Assuming the seeding process is eventually proved, the next tasks include the development of more complex seed molecules such as are needed for selective growth of more complex centers like NE8 [5] or of NV-13C quantum registers. Finally any successful seed growth must be optimized for the growth conditions achievable in large industrial presses. To this end we note that we have grown diamonds at pressures below 5 GPa in some of our diamond anvil experiments.

**Acknowledgements:**

We gratefully acknowledge support of the Carnegie Department of Energy Alliance Center (CDAC), Army Research Laboratory (ORISE program), the National Science Foundation Grants 1202258 and EEC-0540832 (MIRTHE ERC), NIH SBIR contract #HHSN26820150010C, MIT Lincoln Lab PO 7000339119, and the Robert A. Welch Foundation (Award A-1261).

The authors also thank Maddury Somayazulu (Carnegie Institute) for help with resistive heating, Alexander Goncharov (Carnegie Institute) for the use of his laser heating setup, Naira Martirosyan (V.S, Sobolev Institute of Geology and Mineralogy) for assisting with some of the DAC experiments, James Hemmer (UCSB) for suggesting the use 1-adamantylamine to produce NVs. Han-Soo Kim of Texas A&M Univ Microscopy Center of HR-TEM scans of the seeded nanodiamonds, Abdurrahman Almethen, and Masfer Alkahtani of TAMU for assistance with TEM sample preparation.

**Methods (Supplemental Information)**

**1. The diamond anvil cells (DAC) and Multi-Anvil cell**

The precision diagnostics used to monitor nanodiamond growth were made possible by the use of diamond anvil cells (DAC) as growth chambers. As shown in Figure 4(a), DACs have the advantage of being transparent so that optical Raman spectroscopy can be performed before, during and after growth to monitor the chemical state of the growth mixture. In situ X-ray crystallography is also possible. The growth pressures that can be achieved with a DAC are the highest known so that the most extreme growth conditions can be studied. The DAC can also be heated up to 800 $^{o}$C by resistive heating elements, and well above 2000 °C using laser heating as illustrated in Figure 4(a). The disadvantage of a DAC is that the quantities of product are usually very small, ~(50 micron)^3. However using the DAC it is often possible to optimize growth conditions so that much lower pressures (<10 GPa) can ultimately be used. In this case larger HPHT presses can be used to scale up a DAC growth to kilogram quantities.

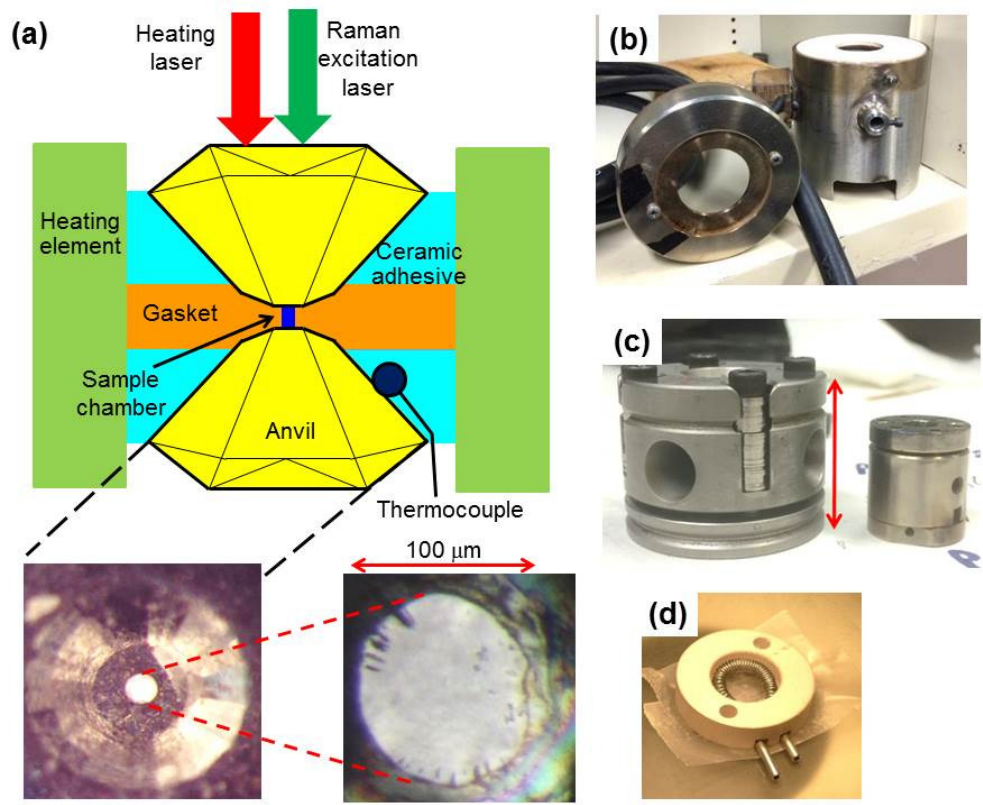

Figure 4. Illustrations and pictures of the DAC experimental equipment. (a) Illustration of the sample chamber inside the DAC used in the resistance heating (furnace) method. In addition to using resistive heating the sample inside the DAC can be locally heated to very high temperatures using a focused heating laser, as illustrated. A key advantage of the DAC is the ability to do real-time diagnostics during growth, for example optical Raman spectroscopy of the reaction mixture. The insets are pictures of the gasket chamber where the sample is contained. Left insert shows the view of the top diamond anvil as seen by the microscope used to record the Raman spectra during growth. The right inset is a close-up of the growth region. (b) Picture of the oven heater, used in place of the furnace design, for the majority of the growth runs. (c) Pictures of the two types of DACs used. The left is the symmetric DAC used with the furnace and laser heating, and the right is the piston cylinder DAC which was used inside the oven. The scale bar is 5 cm. (d) Picture of the Pyrophlite furnace used to heat the sample chamber for the symmetric DAC. The heating wires are covered with a high-temperature epoxy before the experiment begins.

There were two DAC designs used in the experiments, the symmetric DAC and the piston-cylinder (D'Anvils), Figure 4(c). The symmetric DAC was heated using a home-built pyrophlite furnace, Figure 4(d), and the piston-cylinder was heated using an oven (D'Anvils) Figure 4(b). For the furnace heating method the temperature was monitored by gluing a thermocouple onto one of the anvils, which yields an overestimation of the temperature of the sample chamber since the thermocouple is between the heater and the anvils. For the oven, the thermocouple was placed against the outside of the DAC but is expected to give a more accurate reading than the furnace design because of the more uniform temperature inside the oven [28]. For all resistive heating experiments inert gas (98% Ar / 2% $H_2$) was used to prevent oxidation of the DAC and heating elements. Rhenium was used as the gasket material for the symmetric DAC and Inconel was the gasket material for the piston-cylinder DAC. In all experiments the pressure was monitored using the ruby technique [29] [30]. For all experiments the initial diameter of the sample chamber was 100 μm with a depth of ~ 50 μm. The diamond anvil cullet size was 300 μm.

All chemicals were purchased from Sigma-Aldrich. The adamantane was further purified to >99% by zone refinement for all the later experiments (after that of Figure 2). The growth mixture for paraffin seeding experiments was created by melting the paraffin at 60 °C then adding the seed molecule and agitating until it was clearly dissolved in the wax, and for the heptamethylnonane experiments the seed was added inot the liquid and agitated until the seeds were completely dispersed The sample was then placed into the sample chamber by hand using a tungsten needle or a micro-pipette. The mass ratio of the paraffin/ AZADO seeding experiment was ~50/1 and that for paraffin/1-adamantylamine was ~13/1.

A walker-type multi-anvil press was used for the nanodiamond growth experiment pictured in Figure 3(c), located at the Carnegie Institution of Washington. This apparatus comprises a vertical hydraulic ram that applies thrust to a set of 6 pushing anvils contained within a cylindrical retaining ring. The pushing anvils form a central cubic cavity into which the 2nd-stage anvils and sample assembly are placed. The sample assembly is comprised of an octahedron cast from an MgO-based ceramic that contains a cylindrical graphite resistance heater, C-type thermocouple, alumina spacers and the encapsulated starting material. Further details of the pressure module and octahedron manufacture can be found in [31] and references therein. Pressure is applied to the faces of the octahedron via WC cubes with truncated corners (the 2nd-stage anvils). Sample pressure was determined as a function of oil pressure using the e fixed-point calibration curve of [32]. Starting materials were encapsulated using either crushable MgO pieces or graphite.

**2. Comparison of diamondoid seeded and unseeded growth runs**

As stated in the main text the hypothesis of diamondoid seeding can only be supported if no nanodiamonds are formed in the absence of seed molecules. This is verified in Figure 5 which compares two growth runs, one with and one without diamondoid seeds. As seen there are no clear differences in the paraffin C-H Raman spectra during growth. However, without the diamondoid seed molecules no solid product remains after baking the product in air above 530 C (except for large ruby and graphite crystals) as shown by the photos in the inset of Figure 5(c). For the diamondoid seeded case the same high-temperature air baking activates a diamond color center as shown in Figure 5(b). HR-TEM data (not shown) gives independent verification for the presence of diamond only in the diamondoid seeded case. Paraffin-only growth runs were made at temperatures up to 620 C. Note that in all paraffin growth runs with an Inconel gasket graphite rods grew from the gasket, as seen in the inset of see Figure 5(d). This is consistent with the gasket metal catalyzing graphite formation [33]. The growth of these rods was accompanied by a decrease in pressure until reaching a threshold level at which time the graphite rods ceased growing, as shown in Figure 5(d).

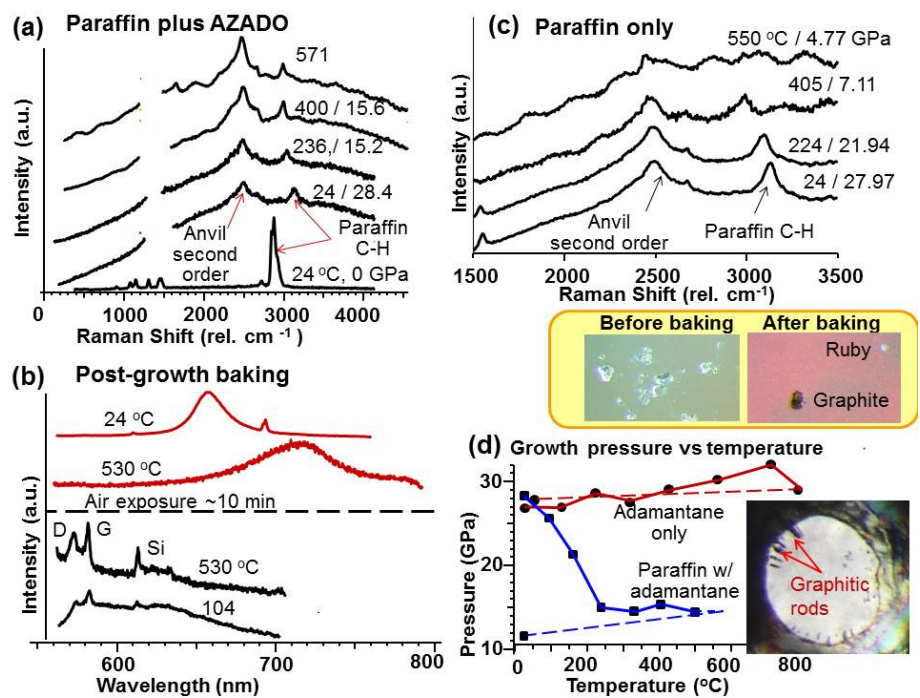

Figure 5. Sensitivity of growth to the presence of diamondoid seed molecules. (a) Raman spectra during diamond growth from paraffin seeded with a modified diamondoid molecule, 2-Azaadamantane *N*-Oxyl (AZADO) (plus an organic phosphorous compound, tricyclohexylphosphine, which was included to provide n-doping to the diamond). Temperatures and pressures are as indicated. Only the C-H stretch of paraffin is visible. (b) Lower two traces show post-growth Raman spectra obtained when the product is baked in a reducing atmosphere of forming gas. As the paraffin residue is evaporated away, the broadband fluorescence disappears and D and G peaks become prominent. From the literature it is known that D can be a mixture of amorphous carbon and nanodiamond and G is crystalline graphite. The upper two traces show that after exposure to air for 10 minutes at 530 C, the D and G peaks almost disappear and fluorescence from a diamond color center appears. The temperature sensitivity of these spectra to temperature is similar to that observed for the color center presented in the bottom trace in Figure 3(d) although the linewidth of the DAC product is narrower. The peak labeled "Si" marks the background from the Si substrate the sample was placed on after extraction from the DAC. (c) Raman spectra obtained for a paraffin-only growth run. Temperatures and pressures are as indicated. Note that the paraffin C-H stretch decreases in amplitude with increasing temperature, just as for the seeded growth. However in this case no solid product remains after exposing to air above 500 C (except for ruby and the graphitic rods that grew from the metal gasket), as shown in the inset photos. This and absence of diamond in HR-TEM images (not shown) confirm that the seed molecules are essential for growing nanodiamond. (d) For all paraffin growth runs the pressure begins dropping as soon as the mixture is heated. This suggests that the paraffin begins cracking at very low temperatures. The inset photo shows graphitic rods that begin growing from the metal gasket as soon as the mixture is heated above ~100 C. When the graphitic rods stop growing the pressure stops decreasing suggesting that a threshold has been reached. In contrast for the adamantane-only growth run of Figure 2 the pressure did not decrease significantly.

### 3. Adamantane-only growth in the oven

Since there was some uncertainty in the exact growth temperature for the furnace-growth run of Figure 2, another adamantane-only growth run was performed using the oven of Figure 4c. The data from this oven-growth run is shown in Figure 6. The adamantane transition at ~450 C in the oven is similar to the one occurring at ~650 C in the furnace (see Figure 2(a)). Some of this temperature difference was due to the overestimation of temperature in the furnace design but since the pressures were very different, no direct comparison can be made. Note that the loss in pressure during growth in the oven design can be attributed to relaxation of the DAC structure which does not occur in the furnace design, or to the fact that the furnace design was implemented with Re gaskets which do not catalyze graphite and therefore consume less adamantane. After growth the pressure in the DAC was increased to its initial value and the resulting Raman spectra appear identical to the initial ones, thus verifying that no significant decomposition of the adamantane occurred.

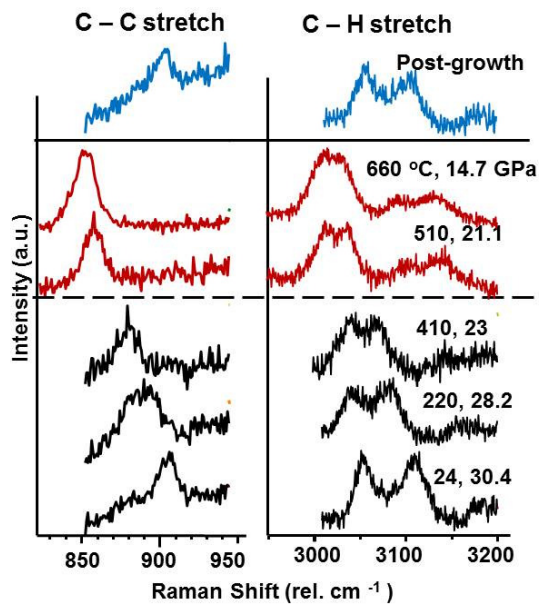

**Figure 6.** Adamantane-only growth run in the oven-heated DAC design of Figure 4b. As before, no significant adamantane decomposition is observed. Raman spectra were taken during growth at the temperatures and pressures indicated. As the temperature begins to increase both the C – C and C – H stretch frequencies gradually decrease by an amount that is consistent with the decreasing pressure. Above 410 C (horizontal dashed line) there is a shift toward smaller wavenumbers for both spectra which suggests that the adamantane has undergone a transformation. Possibly this is a phase transition since once the growth run is completed and the temperature and pressure are restored to their initial values both bands return to their initial positions and linewidths (post-growth spectra above the solid line).

## References

1. Wentorf, R.H., *The Behavior of Some Carbonaceous Materials at Very High.* TheJournal of Physicsl Chemistry, 1965: p. 3063-3069.
2. Davydov, V.A., et al., *Production of nano- and microdiamonds with Si-V and N-V luminescent centers at high pressures in systems based on mixtures of hydrocarbon and fluorocarbon compounds.* Jetp Letters, 2014. **99**(10): p. 585-589.
3. Hong, S., *Nanoscale magnetometry with NV centers in diamond.* MRS Bulletin, 2013: p. 155-161.
4. Vlasov, I.I., A.A. Shiryaev, and T. Rendler, *Molecular-sized fluorescent nanodiamonds.* Nature Nanotechnology, 2014. **9**: p. 54.
5. Rabeau, J.R., et al., *Fabrication of single nickel-nitrogen defects in diamond by chemical vapor deposition.* Applied Physics Letters, 2005. **86**(13).
6. Davydov, V.A., *Nanosized carbon forms in the process of pressure-temperature-induced transformations of hydrocarbons.* Carbon, 2006. **44**: p. 2015-2020.
7. Davydov, V.A., et al., *Synergistic Effect of Fluorine and Hydrogen on Processes of Graphite and Diamond Formation from Fluorographite-Naphthalene Mixtures at High Pressures.* Journal of Physical Chemistry C, 2011. **115**(43): p. 21000-21008.
8. Ishiwata, H., et al. *Fluorescent Nanodiamonds from Molecular Diamond Seed*. in *CLEO: 2015*. 2015. San Jose, California: Optical Society of America.

9. Dahl, J.E.P., et al., *Synthesis of Higher Diamondoids and Implications for Their Formation in Petroleum.* Angewandte Chemie-International Edition, 2010. **49**(51): p. 9881-9885.
10. Shizuno, T., et al., *Synthesis of Diamondoids by Supercritical Xenon Discharge Plasma.* Japanese Journal of Applied Physics, 2011. **50**(3).
11. Maurer, P.C., et al., *Room-Temperature Quantum Bit Memory Exceeding One Second.* Science, 2012. **336**(6086): p. 1283-1286.
12. Lesik, M., et al., *Perfect preferential orientation of nitrogen-vacancy defects in a synthetic diamond sample.* Applied Physics Letters, 2014. **104**(11).
13. Fukui, T., et al., *Perfect selective alignment of nitrogen-vacancy centers in diamond.* Applied Physics Express, 2014. **7**(5).
14. Michl, J., et al., *Perfect alignment and preferential orientation of nitrogen-vacancy centers during chemical vapor deposition diamond growth on (111) surfaces.* Applied Physics Letters, 2014. **104**(10).
15. Pinault-Thaury, M.A., et al., *n-Type CVD diamond: Epitaxy and doping.* Materials Science and Engineering B-Advanced Functional Solid-State Materials, 2011. **176**(17): p. 1401-1408.
16. Mochalin, V.N., et al., *The properties and applications of nanodiamonds.* Nature Nanotechnology, 2012. **7**(1): p. 11-23.
17. Jelezko, F., et al., *Observation of coherent oscillations in a single electron spin.* Physical Review Letters, 2004. **92**(7).
18. Dutt, M.V.G., et al., *Quantum register based on individual electronic and nuclear spin qubits in diamond.* Science, 2007. **316**(5829): p. 1312-1316.
19. Fallas, J.C., et al., *Raman Spectroscopy Measurements of the Pressure-Temperature Behavior of LiAlH4.* Journal of Physical Chemistry C, 2010. **114**(27): p. 11991-11997.
20. Vijayakumar, V., et al., *Pressure induced phase transitions and equation of state of adamantane.* Journal of Physics-Condensed Matter, 2001. **13**(9): p. 1961-1972.
21. Saxena, S.K., H.P. Liermann, and G.Y. Shen, *Formation of iron hydride and high-magnetite at high pressure and temperature.* Physics of the Earth and Planetary Interiors, 2004. **146**(1-2): p. 313-317.
22. Bragg, L.B., *Rate of cracking of paraffin wax.* Industrial and Engineering Chemistry, 1941. **33**: p. 376-380.
23. Ferrari, A.C. and J. Robertson, *Raman spectroscopy of amorphous, nanostructured, diamond-like carbon, and nanodiamond.* Philosophical Transactions of the Royal Society a-Mathematical Physical and Engineering Sciences, 2004. **362**(1824): p. 2477-2512.
24. Gaebel, T., et al., *Size-reduction of nanodiamonds via air oxidation.* Diamond and Related Materials, 2012. **21**: p. 28-32.
25. Bennett, N.R., J.M. Brenan, and Y.W. Fei, *Metal-silicate Partitioning at High Pressure and Temperature: Experimental Methods and a Protocol to Suppress Highly Siderophile Element Inclusions.* Jove-Journal of Visualized Experiments, 2015(100).
26. Dolde, F., et al., *Electric-field sensing using single diamond spins.* Nature Physics, 2011. **7**(6): p. 459-463.
27. Neumann, P., et al., *High-Precision Nanoscale Temperature Sensing Using Single Defects in Diamond.* Nano Letters, 2013. **13**(6): p. 2738-2742.
28. Munro, R.G., et al., *TEMPERATURE DISTRIBUTION IN THE DIAMOND ANVIL PRESSURE CELL AT HIGH-TEMPERATURE.* Journal of Applied Physics, 1984. **55**(1): p. 4-8.
29. Mao, H.K., J. Xu, and P.M. Bell, *CALIBRATION OF THE RUBY PRESSURE GAUGE TO 800-KBAR UNDER QUASI-HYDROSTATIC CONDITIONS.* Journal of Geophysical Research-Solid Earth and Planets, 1986. **91**(B5): p. 4673-4676.

30. Rekhi, S., L.S. Dubrovinsky, and S.K. Saxena, *Temperature-induced ruby fluorescence shifts up to a pressure of 15 GPa in an externally heated diamond anvil cell.* High Temperatures-High Pressures, 1999. **31**(3): p. 299-305.
31. Walker, D., *LUBRICATION, GASKETING, AND PRECISION IN MULTIANVIL EXPERIMENTS.* American Mineralogist, 1991. **76**(7-8): p. 1092-1100.
32. Bertka, C.M. and Y.W. Fei, *Mineralogy of the Martian interior up to core-mantle boundary pressures.* Journal of Geophysical Research-Solid Earth, 1997. **102**(B3): p. 5251-5264.
33. Giardini, A.A. and J.E. Tydings, *DIAMOND SYNTHESIS - OBSERVATIONS ON MECHANISM OF FORMATION.* American Mineralogist, 1962. **47**(11-1): p. 1393-&.